\documentclass[aps,prd,superscriptaddress,nofootinbib,twocolumn]{revtex4} 
\usepackage{graphics}
\usepackage{epsfig}
\usepackage[english]{babel}
\usepackage{amsmath} % Useful for equation alignment
\usepackage{verbatim} % Allows to have comments over multiple lines
\usepackage{mathrsfs}
\usepackage{amsfonts}
\usepackage{amssymb}
\usepackage[latin1]{inputenc}
\usepackage{hyperref}

\newcommand{\beq}{\begin{equation}}
\newcommand{\eeq}{\end{equation}}
\newcommand{\vk}{\textbf{k}}
\newcommand{\dphi}{\delta \phi}
\newcommand{\x}{\textbf{x}}

\newcommand{\bra}{\langle}
\newcommand{\ket}{\rangle}
\newcommand{\mH}{\mathcal{H}}
\newcommand{\mP}{\mathcal{P}}

\newcommand{\barr}{\begin{eqnarray}}
\newcommand{\earr}{\end{eqnarray}}
\newcommand{\bea}{\begin{eqnarray*}}
\newcommand{\eea}{\end{eqnarray*}}
\newcommand{\nn}{\nonumber \\}

  \newcommand{\eq}[1]{Eq. \eqref{#1}}
 
 \newcommand{\RI}{\textrm{R,I}} 
 
   \newcommand{\avg}[1]{\mathbb{E} \{ {#1} \}}
  \newcommand{\expec}[1]{\langle {#1} \rangle}

\begin{document}

%\title{Eternal inflation and the dynamical reduction of the wave function}
\title{Eternal inflation and the quantum birth of cosmic structure}

\author{Gabriel Le\'{o}n}
\email{gleon@fcaglp.unlp.edu.ar}
\affiliation{Grupo de Astrof\'isica, Relatividad y 
Cosmolog\'ia, Facultad de Ciencias 
Astron\'omicas y Geof\'isicas, Universidad Nacional de La Plata, Paseo del Bosque S/N 
(1900) La Plata, Argentina}
\affiliation{Departamento de F\'isica, Facultad de Ciencias Exactas y Naturales, Universidad de Buenos Aires, Ciudad Universitaria -
  Pab. I, 1428 Buenos Aires, Argentina.}

\begin{abstract}

  We consider the eternal inflation scenario of the slow-roll/chaotic type with the additional element of an objective collapse of the wave function. The incorporation of this new agent to the traditional inflationary setting might represent a possible solution to the quantum measurement problem during inflation, a subject that has not reached a consensus among the community. Specifically, it could provide  an explanation for the generation of the primordial anisotropies and inhomogeneities, starting from a perfectly symmetric background and invoking symmetric dynamics. We adopt the continuous spontaneous localization model, in the context of inflation, as the dynamical reduction mechanism that generates the primordial inhomogeneities. Furthermore, when enforcing the objective reduction mechanism, the condition for eternal inflation can be bypassed.  In particular, the collapse mechanism incites the wave function, corresponding to the inflaton, to localize itself around the zero mode of the field.  Then the zero mode will evolve essentially unperturbed, driving inflation to an end in any region of the Universe where inflation occurred. Also, our approach achieves a primordial spectrum with an amplitude and shape consistent with the one that best fits the observational data.

\end{abstract}

%\keywords{quantum collapse, CMB, cosmology}
%\pacs{98.80.Cq,98.70.Vc,98.80.-k}

\maketitle

\section{Introduction}
\label{intro}

The theory of the early Universe enjoying the most recognition amongst cosmologists is inflation \cite{starobinsky,guth,linde,albrecht}.  The inflationary epoch is characterized by an accelerating expansion of the Universe, and its current success is based on  the power to explain the primordial inhomogeneities generation that represent the seeds of cosmic structure \cite{mukhanov81,mukhanov2,starobinsky2,hawking,hawking2}. Moreover, the recent\textit{ Planck} satellite data release reveals that the description of the primeval  Universe is consistent with the inflationary paradigm \cite{planck2015,planck2015likelihoods,planck2015inflation}. These data suggest that the primordial perturbations spectrum is very close to scale invariance, favoring the simplest inflationary models \cite{planck2015inflation,Martin14}. 

In spite of the  {strong agreement between the inflationary predictions and the observational data, there are still some unsettled conceptual issues; among these is the usually called quantum-to-classical transition of the primordial perturbations.}\footnote{ {See for instance Sec. 10.1 of \cite{weinberg2008} and the end of Sec. 8.3 of \cite{mukhanovbook}.  }}   {As initially presented in \cite{PSS}, a precise formulation of that issue is as follows: }
%the standard account lacks a crucial element capable of addressing the precise physical mechanism for generating the actual primordial inhomogeneities. Specifically, the traditional inflationary picture, as initially argued in \cite{PSS}, and extensively exposed in \cite{shortcomings,alberto,susana2013}, does not provide a detailed explanation on how 
Starting from an initial situation which is taken to be perfectly isotropic and homogeneous,  {both in the background spacetime and the quantum state characterizing the matter fields,} and based on a dynamics that supposedly preserves those symmetries, one ends with a non-symmetric situation corresponding to the inhomogeneities and anisotropies observed  in the Universe. 

 {Most of the accepted proposals to address the quantum-to-classical transition of the perturbations during inflation are based on the decoherence framework \cite{halliwell,grishchuk,polarski,polarski1996,kiefer1998,kiefer2,kiefer2006,kiefer}, with a varying degree of reliance on non-orthodox interpretations of quantum mechanics, e.g. many-worlds \cite{nomura2011,carroll}, and the consistent histories formulation \cite{hartle1992} (although we do not subscribe to such proposals because of the arguments exposed in \cite{shortcomings,susana2013,elias2015}). In fact, in recent works \cite{markkanen2016,markkanen2017}, it has been shown how decoherence in de Sitter spacetime can lead to a violation of the time translation invariance (nonetheless, the spatial and vacuum symmetries remain intact). In any regard it is fair to say that the quantum measurement problem, and in particular in the cosmological context, is still very much an unsolved issue. Therefore, it is reasonable to explore all consistent proposed solutions.}  One possible solution to the aforementioned shortcoming is to consider an objective dynamical reduction mechanism\footnote{For other approaches addressing the shortcomings of the quantum measurement problem during inflation see Refs. \cite{pintoneto,valentini,magueijo2016}  .} \cite{PSS,shortcomings,susana2013,micol}. Such a process can break the homogeneity and isotropy of the inflaton vacuum state and, in turn,  generates the primordial curvature perturbation.

In the present work, we employ the continuous spontaneous localization (CSL) model as the objective reduction mechanism. The CSL model belongs to a large class of models simply called \textit{collapse models}, which attempt to provide a solution to the measurement problem of Quantum Mechanics \cite{GRW,diosi1989,pearle,pearle2,bassi}.  The common idea in these models is to modify the Schr\"odinger equation by introducing nonlinear stochastic corrections that spoil its linearity, inducing a spontaneous random localization of the wave function in a sufficiently small region of space. The model parameters are chosen in a way that ensures that micro-systems evolve following closely the dynamics provided by the Schr\"odinger equation; conversely, macro-systems are extremely sensible to the nonlinear effects resulting in a sharply localized wave function. Nevertheless, the main aspect of the CSL model is that there is no need to mention or to introduce the notion of an observer/measurement device as in the orthodox interpretation. This is a desired  element in the context of the early Universe and cosmology in general. The implementation of the CSL model into the inflationary scenario has been analyzed in previous works \cite{pedro,jmartin,hinduesS}. Some of them resulting in theoretical predictions consistent with the observational data, in particular, with the observed amplitude and shape of the primordial spectrum (scalar and tensor) \cite{pedro,hinduesT,LB15,qmatter,abhishek}.

 {Our view to implement the CSL model into inflation relies on the semiclassical gravity approximation, described by Einstein equations $ $  $G_{ab} = 8 \pi G \bra \hat T_{ab} \ket$. In this approximation, the matter degrees of freedom are treated quantum mechanically, while the gravitational sector, characterized by the spacetime metric is always classical. The validity of the semiclassical gravity approach is limited. In particular, it requires that the scalar curvature of the spacetime to be small compared to $l_P^{-2}$ ($l_P$ is the Planck length). It also requires that the fluctuations in the energy-momentum tensor are small compared to its absolute value \cite{Kuo1993}. The inflationary epoch is assumed to be associated with energy scales that are way below the Planck regime. Regarding the second constraint, we expect that once the collapse mechanism has ended, the quantum fluctuations of the energy--momentum tensor should be small, although we cannot be completely sure until a full covariant reduction mechanism is developed (see Ref. \cite{Tumulka2005,Bedingham2010,Pearle2015,benito} for recent advances in that direction). }

In this paper, we are concerned with the subject of eternal inflation \cite{linde1982,linde1986,linde1986b,vilenkin1982,vilenkin1983,goncharov1987,starobinsky1982}.  {In particular, we will  focus only on the eternal inflation mechanism provided by potentials of the slow-roll/chaotic type \cite{guth2007}. At this point, it is worthwhile to differentiate the  eternal chaotic inflation from the one produced by false vacuum tunneling \cite{vilenkin1983,guth2007}; the analysis of the latter  type of inflationary models will be left for future work. Therefore, unless we state it explicitly, when referring to eternal inflation throughout this work,  we have in mind the eternal chaotic inflation  kind.   }

The traditional accepted idea,  {regarding eternal inflation,} is to consider the vacuum quantum uncertainty of the inflaton  as a true dynamical event, or commonly called \textit{quantum fluctuations}. In the textbook presentation of inflation, the classical trajectory of the field is governed by the dynamics of the zero mode. Inflation unfolds as the zero mode slowly rolls down the potential, ending when the field lies at the bottom of the potential.   However, the zero mode can \emph{fluctuate} due to  the quantum fluctuations of the inflaton.\footnote{Although, strictly speaking, we point out that there are no fluctuations, only quantum uncertainties.} The fluctuations can be sufficiently large that the field spontaneously goes upward the potential, starting a new cycle of inflation. This behavior can arise in any region of the Universe, thus, achieving a self-reproducing state of inflation, ordinarily called eternal inflation. It is known that eternal inflation  is a typical feature of  inflationary models with potentials that best fit the observational data \cite{kinney2}. 

Perhaps one of the most striking consequences of eternal inflation is its connection with a multiverse \cite{vilenkin1983,guth2000,guth2013}, in which ``anything that can happen will happen, and it will happen an infinite number of times'' \cite{guth2007}. For instance, the constants of nature can take a wide range of values depending on the region of the multiverse \cite{vilenkin2011,sergei}.  {It is important to remark that the connection between eternal inflation and the multiverse is more rigorous in the case of eternal inflation via tunneling between false vacua than the eternal inflation mechanism considered in the present work. } Some cosmologists view the multiverse as an issue questioning the validity of the inflationary theory \cite{ijjas,ijjas2,turok}, while others consider the multiverse hypothesis admissible and positive  \cite{linde3,vilenkin2011,guth2013,guth4}. Nevertheless, as the multiverse is a widely accepted consequence of eternal inflation, we think it is worthwhile to meticulously analyze all the underlying assumptions of the possible eternal inflation scenarios (see Refs. \cite{Feng2010,Qiu2011} for other related works).

One of the basic premises of eternal inflation is that it is appropriate to treat the vacuum quantum uncertainty of the inflaton as an actual dynamical object, which literally disturbs the homogeneous part of the field, i.e. the zero mode. We find this picture misleading. For example, consider the one-dimensional quantum harmonic oscillator in its ground state. Ehrenfest's theorem guarantees that the expectation value of the momentum and position operators will follow the same dynamics as its classical counterparts. Nonetheless, there is nothing in ordinary quantum mechanics suggesting the need to modify the classical motion as a result of the quantum uncertainties. Also, the postulates of quantum mechanics indicate that, after a measurement of, say, the position, the amplitude of the wave function  will change into a wave packet narrowly centered around the measured value. The peak of the wave packet, i.e. the expectation value, will continue to evolve according to the oscillator classical equations of motion. Moreover,  before the measurement, the quantum state of the harmonic oscillator possessed a reflection symmetry. The dynamics of the vacuum quantum uncertainty does not lead to a breakdown of said symmetry.  Only after the measurement process, the reflection symmetry of the wave function is lost.

In the case of the inflaton $\hat \phi(\x,t)$, the vacuum expectation value is precisely equal to the zero mode $ $ $\bra \hat \phi (\x,t) \ket = \phi_0(t)$ (the expectation value of the inhomogeneous part $\bra \hat \dphi (\x,t) \ket$ is zero). Therefore, the evolution of the field expectation value follows the zero mode equation of motion. Although, as in the case of the harmonic oscillator, there is nothing in the postulates of the quantum theory that signals a modification of the classical trajectory of the field; in particular, not an alteration that includes the quantum uncertainty. 
%Nevertheless, the majority of cosmologists seem to favor the decoherence framework \cite{kiefer,halliwell,kiefer2,polarski,grishchuk} with a varying degree of reliance on non-orthodox interpretations of quantum mechanics, e.g. many-worlds \cite{nomura2011,carroll}. However, as it has been argued in previous works \cite{shortcomings,susana2013,elias2015}, neither decoherence nor the many-worlds interpretation offers a completely satisfactory solution for the problem we have indicated, i.e. the transition from an initial symmetric state into a non-symmetric one, given that its dynamics preserve the original symmetry. 
Furthermore, even if somehow one incorporates the quantum uncertainty into the inflaton dynamics, it is not clear that  $\bra \hat \dphi(\x,t)^2 \ket^{1/2}$, taken in the vacuum state, should be associated to a classical stochastic field, which alters the evolution of the zero mode. At most it can be associated to the amplitude of a stochastic field, but the random nature of its phase will remain unknown. Another issue is to identify the element that plays the role of the measurement device, in the inflationary Universe, as in an ordinary laboratory situation.  {Thus,  eternal inflation is deeply linked with the quantum-to-classical transition and the quantum measurement problem in general.  }

On the other hand, incorporating the CSL model into inflation could help to clarify the aforementioned issues  {(although open questions will remain since at the present there is no formal version of the CSL model for quantum fields)}. In particular, it can successfully change the symmetries of the vacuum state and, at the same time, be responsible for the birth of the primordial curvature perturbation \cite{pedro,elias2013}. In addition, since the evolution provided by the CSL model transforms spontaneously the vacuum state, it is possible to formally identify the expectation value $\bra \hat \dphi(\x,t) \ket$ with a classical stochastic field $\dphi(\x,t)$.  The localization of the quantum state of the field, due to the CSL dynamics, implies that the quantum uncertainty of the field is decreasing during inflation, and becomes centered around a value that coincides with $\phi(\x,t) = \phi_0(\eta) + \dphi(\x,t)$, which in turn avoids the condition for eternal inflation. Moreover, given that $\dphi \ll \phi_0$, in the usual interpretation, the zero mode does not have very large fluctuations.

% 
% HASTA AQUI
% 
% 
% --INFLACION NORMALITA Y EXITOS \checkmark \\
% 
% --EL PROBLEMA FUNDAMENTAL Y NUESTRA PROPUESTA DE COLAPSO \checkmark \\
% 
% --CSL, IDEA DEL MODELO Y LOS RESULTADOS OBTENIDOS HASTA AHORA EN INFLACION \checkmark \\
% 
% --ETERNAL INFLATION, VENTAJA O DESVENTAJA? MULTIVERSOS \checkmark\\
% 
% --COSAS ESOTERICAS DE CUANTICA Y SU RELACION CON ETERNAL INFLATION\\
% 
% --NUESTRA IDEA DE COMO EL COLAPSO MODIFICARIA LO QUE TIENE QUE VER CON ETERNAL INFLATION\\

 The paper is structured as follows:
In Sec. \ref{standard}, we present a brief review of slow-roll inflation focusing on the main aspects leading to the eternal inflation picture. Then, in Sec. \ref{secCSL}, we focus on the important features of the CSL model concerning its implementation into the inflationary context. Afterwards, in Sec. \ref{mod}, we show how the CSL inflationary model modifies the usual conclusions drawn from slow-roll inflation resulting in the eternal inflation scenario. Finally, Sec. \ref{disc} contains discussion and conclusions.

\section{The standard picture}
\label{standard}

We will begin by reviewing the standard picture of eternal inflation within the framework 
of slow-roll inflation \cite{sergei,guth2013}. 

The inflationary Universe is described by Einstein equations 
$G_{ab} = 8 \pi GT_{ab}$ ($c=1$). In slow-roll inflation, the matter sector is 
characterized by a single scalar field, the inflaton $\phi$. Meanwhile, the background 
spacetime is described by a FRW Universe with line element (in conformal 
coordinates) $ds^2 = a^2(\eta) [-d\eta^2 + \delta_{ij} dx^i dx^j]$. The scale factor can 
be approximated by $a(\eta) \simeq -1/[H\eta(1-\epsilon_1)]$, with $H$ the Hubble factor, 
which during 
inflation is approximately constant and $\eta$ the conformal time $\eta \in (-\infty,0)$. 
The deviation from a perfect de Sitter expansion is parameterized by the first slow 
parameter defined as $ \epsilon_1 \equiv  1 - {\mH'}/{\mH^2} $, where  $\mH \equiv a'/a = 
aH$ is the conformal expansion rate and the prime denotes derivative with respect to 
$\eta$.  The energy density of the Universe is dominated by the 
 inflaton potential $V(\phi)$, and during slow-roll inflation 
 $ \epsilon_1 \simeq  (M_P^2/2) (\partial_\phi V/V)^2 \ll 1$, with
$M_P^2 \equiv (8\pi G)^{-1}$ the reduced Planck mass.

The action of the inflaton with mass $m$ minimally coupled to  gravity leads to a 
quantum field theory for $\phi$ in a quasi-de Sitter spacetime. In particular, one can 
decompose the quantum field in Fourier modes
\begin{equation}\label{campototal}
  \hat \phi(\x,\eta) = \frac{1}{L^3} \sum_{\vk} \hat \phi_{\vk}(\eta) e^{i \vk 
\cdot \x}.
\end{equation} 
In order to avoid infrared divergences, we have introduced a regularization and consider 
the field in a box of side $L$. The sum is over the wave vectors $\vk$ satisfying $k_i 
L = 2 \pi n_i$ for $i = 1, 2, 3$ with $n_i$ integers, and the field operator operator given by 
\begin{equation}
  \hat \phi_{\vk} (\eta) = \phi_k(\eta) \hat a_{\vk} + \phi_k^* (\eta) \hat 
a^{\dag}_{-\vk}.
\end{equation} 
The mode functions $\phi_k$ satisfy the evolution equation
\begin{equation}\label{kleingordon}
  \phi_{k}'' + 2 \mH \phi_k' + (k^2 + a^2m^2) \phi_k = 0
\end{equation} 
and the normalization condition
\begin{equation}\label{simplectic}
  \phi_k \phi^{*'}_k - \phi_k' \phi_k^* = i a^{-2}.
\end{equation} 
It is known that the normalization condition \eqref{simplectic} does not determine the set of mode solutions 
unequivocally; a particular choice of the solutions of \eq{kleingordon} implies an 
election of the vacuum. For $k=0$, the general solution to \eq{kleingordon} is a linear 
combination of $\eta^{(3+2\epsilon_1-2\nu)/2}$ and $\eta^{(3+2\epsilon_1+2\nu)/2}$, with 
$\nu = 3/2+\epsilon_1-m^2/(3H^2)$. For $k \neq 0$, the general solution is a linear 
combination of $\eta^{3/2+\epsilon_1} H_{\nu}^{(1)} (-k\eta)$ and $\eta^{3/2+\epsilon_1} 
H_{\nu}^{(2)} (-k\eta)$, with $H_\nu^{(1)}$ and $H_\nu^{(2)}$ the Hankel functions of 
first and second kind of order $\nu$. We now choose the Bunch-Davies vacuum, i.e. we 
choose the modes such that at earlier times $\eta \to -\infty$ they behave as ``positive 
frequency solutions'', normalized according to \eq{simplectic}.  {In fact, we highlight that the Bunch-Davies vacuum in an expanding space is more than simply a choice: for any given initial condition with certain quite natural assumptions for the UV behavior, the Bunch-Davies vacuum is always approached as it is a late-time attractor solution, see Refs; \cite{anderson2000,markkanen2017b}.  } Therefore the 
Bunch-Davies vacuum corresponds to select the functions $H_\nu^{(1)}$. The operators $\hat a_{\vk}$ and $\hat a_{\vk}^{\dag}$ are the 
annihilation and creation operators respectively, which satisfy the commutation rules 
$[\hat 
a_{\vk},\hat a_{\vk'}^{\dag}] = \delta_{\vk,\vk'}$. As usual the vacuum is defined by 
$\hat a_{\vk} |0 \ket = 0$ for all $\vk$. 

The inflaton can be split into an homogeneous part denoted by $\hat \phi_0$ plus  
small inhomogeneities, i.e.
\begin{equation}
  \hat \phi(\x,\eta)  = \hat \phi_0 (\eta) + \varepsilon \hat \dphi (\x,\eta),
\end{equation} 
where $\varepsilon \ll 1$ is a parameter (do not confuse with the first slow-roll 
parameter $\epsilon_1$) that helps to quantify the smallness of the inhomogeneities.
The homogeneous part $\hat \phi_0 (\eta) $ corresponds to the zero mode of 
the field. Therefore, we can rewrite \eq{campototal} as follows
\begin{equation}
  \hat \phi(\x,\eta) =  
\frac{\hat \phi_{\vk=0} (\eta)}{L^3} + \frac{\varepsilon}{L^3} \sum_{\vk\neq 0} \hat 
\dphi_{\vk}(\eta) e^{i \vk 
\cdot \x},
\end{equation} 
where $\hat \dphi_{\vk}$ denotes the modes of the field $\hat \phi_{\vk}$ with $k \neq 0$.

The initial state of the field $\hat \phi(\x,\eta)$ is a featureless vacuum state, 
except for the zero mode that is excited. Explicitly the initial state is
\begin{equation}\label{edoini}
  | \textrm{in} \ket = |\xi_0 \ket \bigotimes_{\vk} | 0 \ket_{\vk}, 
\end{equation} 
with $\vk \neq 0$.  For $\vk \neq 0$, the state is the Bunch Davies vacuum. For  the zero mode we assume a 
coherent state, i.e. a state such that $\hat a_0 |\xi_0 \ket = \xi_0 |\xi_0 \ket $. The assumption that $| \xi_0 \ket$ corresponds to a coherent state is justified by noting  that  such states are sharply 
peaked around a value that coincides with the homogeneous part of the classical field 
$\phi_0(\eta)$. In other words, the state $|\xi_0 \ket$ satisfies $\bra \xi_0 
|\hat \phi_{\vk=0} (\eta) |\xi_0 \ket = \phi_0(\eta)(\xi_0+\xi_0^*)$. Consequently, 
\begin{equation}
 \bra \textrm{in} | \hat \phi (\x,\eta)  | \textrm{in} \ket \propto
 \phi_0(\eta).
\end{equation}
Additionally,  expectation values satisfy their counterpart classical equations of motion. In particular, the homogeneous classical field $\phi_0(\eta)$ satisfies
\begin{equation}\label{motionclassical}
 \phi_0''(\eta) +  2 \mH \phi_0' (\eta) +  a^2 m^2 \phi_0(\eta) = 0,
\end{equation} 
which is the well known equation of motion corresponding to the (classical) 
homogeneous part of the inflaton. We now recall the definition of the second slow-roll 
parameter, $\epsilon_2 \equiv \epsilon_1'/\mH \epsilon_1$. In terms of the slow-roll 
parameters, \eq{motionclassical} is
\begin{equation}
 3\mH\phi_0'(\eta)\left[ 1-\frac{\epsilon_1}{3} + \frac{\epsilon_2}{6}   \right] = -a^2 
m^2 \phi_0(\eta).
\end{equation} 
Assuming the slow-roll approximation $\epsilon_1 \ll 1$ and $\epsilon_2 \ll 1$, we 
recover the usual slow-roll equation for the homogeneous field $3 \mH \phi_0' \simeq -a^2 
m^2 \phi_0$.

On the other hand, the quantum uncertainty\footnote{In fact, the definition given in Eq. \eqref{uncert} is properly  the quantum uncertainty squared. The quantum uncertainty of a quantum operator $\hat X$ is defined as $\delta_Q \hat X \equiv \sqrt{ \bra \hat X^2 \ket - \bra \hat X \ket^2}$; however, we will abuse the language and refer to both $\delta_Q^2 \hat X$ and $\delta_Q \hat X$ as the quantum uncertainty.    }      of the field $\hat \phi(\x,\eta)$,
defined as 
\begin{equation}\label{uncert}
\delta_Q^2 \hat \phi (\x,\eta) \equiv  \bra \textrm{in} | \hat \phi (\x,\eta)^2 | \textrm{in} \ket - (\bra \textrm{in} | \hat \phi (\x,\eta)  |  \textrm{in} \ket)^2, 
\end{equation} 
   is not zero. In particular, it is given by
\begin{eqnarray}
\delta_Q^2 \hat \phi (\x,\eta) & =&     \frac{ \bra \xi_0 | \hat \phi_{0}  
  (\eta)^2 | \xi_0 \ket }{L^6}   - \frac{(\phi_0(\eta) 2 \textrm{Re} [\xi_0])^2}{L^6} \nn
&+&\frac{ \varepsilon^2}{L^6} \sum_{\vk,\vk' \neq 0} \bra 0 
| \hat \dphi_{\vk} (\eta) \hat \dphi_{\vk'} (\eta) | 0 \ket e^{i (\vk + \vk') \cdot \x}.\nn
\end{eqnarray} 
If we assume a renormalized energy-momentum tensor, then we can consider normal 
ordering. Therefore, for the coherent state $| \xi_0 \ket$, one has $  \bra \xi_0 |  
{\hat \phi_{0}  (\eta)^2} | \xi_0 \ket = {(\phi_0(\eta) 2 \textrm{Re} 
[\xi_0])^2}$, i.e. the quantum uncertainty of the zero mode vanishes. Thus, the quantum uncertainty of the field $\hat \phi$ is 
\begin{equation}
  \delta_Q^2 \hat \phi (\x,\eta) = \varepsilon^2  \bra 0 | \hat \dphi (\x,\eta)^2 | 0 \ket .
 \end{equation}

In summary, the quantum field $\hat \phi (\x,\eta)$ is in the initial state $ | 
\textrm{in} \ket$. The expectation value $  \bra \hat \phi (\x,\eta)  \ket $ is 
proportional to the homogeneous (classical) field 
$\phi_0$. The quantum uncertainty of the field $\hat \phi (\x,\eta)$  is given 
by ${\bra 0 | \hat \dphi (\x,\eta)^2 | 0 \ket}$, that is, by the quantum uncertainty of 
the inhomogeneous part characterizing the inflaton.

In the traditional picture, eternal inflation is achieved when the variation of the quantum uncertainty of the field becomes large enough, so  that the value of the field $\phi$ cannot be properly localized. In other words, the standard approach considers the quantum uncertainty  as an actual dynamical fluctuation, referred to as \textit{quantum fluctuations.} These quantum fluctuations will be  superimposed onto the deterministic slow-roll trajectory obeyed by $\phi_0$. The fluctuations will lead the value of $\phi$ away from $\phi_0$ in some regions, originating a new cycle of inflation in those regions. This process will occur endlessly, resulting in an infinite number of regions inflating, sometimes called ``pocket universes'' \cite{guth2007}. In some regions inflation may end, while others continue inflating.

Formally,  eternal inflation arises when the variation of  $\phi_0$  over one Hubble time is smaller than the  variation of $\bra 0 | \hat \dphi (\x,\eta)^2 | 0 \ket$  over the same  time period.  The mathematical expression can be deduced as follows.

The variation of $\phi_0$ over one Hubble time is simply
\begin{equation}\label{varphi}
  \Delta \phi_0(t) = \dot \phi_0 (t) \Delta t = \dot \phi_0(t) H^{-1},
\end{equation} 
where we have switched from conformal to cosmic time $t$ (the dot over the function represents derivative with respect to $t$).

Next, one needs an estimate for the amplitude associated with the variation of the quantum uncertainty ${\bra 0 | \hat \dphi (\x,\eta)^2 | 0 \ket}$ over one Hubble time. That estimate is usually  obtained by introducing an UV cut-off \cite{brandenberger}    $k_H(\eta)$, which is the comoving wave number $k_H$ corresponding to Hubble radius crossing at a time $\eta$, so $k_H(\eta) = -1/\eta$. Physically, the cut-off implies one is considering only super-Hubble modes since these modes  contribute  most to the quantum uncertainty amplitude.  Therefore, the variation of the quantum fluctuations, in terms of the Fourier 
modes with $k \neq 0$, is
\begin{eqnarray}\label{incert1}
  \Delta [ \delta_Q^2 \hat \phi (\x,t)]&=& \Delta t  \frac{1}{2\pi^2}  
  \frac{d\eta}{dt} \frac{d}{d\eta}\bigg[ \int_{0^+}^{k_H (\eta)} 
  {dk}\:k^2 |\dphi_k|^2 \bigg] \nn
  &=& \Delta t  \frac{1 }{ 2 \pi^2} a(\eta)^{-1} |\dphi_{k_H}|^2 k_H^2 \frac{d k_H}{d 
    \eta} .
\end{eqnarray} 
Approximating the mode functions $\dphi_k \sim \eta^{3/2} H_{3/2}^{(1)}$, and $a(\eta) \simeq -1/(H \eta)$, \eq{incert1} yields
\begin{equation}\label{varphiquant}
  \Delta [ \delta_Q^2 \hat \phi (\x,t)]\simeq \frac{\Delta t H^{3}}{4 \pi^2} = 
  \frac{H^{2}}{4 \pi^2}.
\end{equation} 

At this point, we can provide the condition for eternal inflation, using Eqs. \eqref{varphi} and \eqref{varphiquant}, the explicit condition is \cite{guth2007,linde3} 
\begin{equation}\label{condicionchingona0}
\frac{ \sqrt{ \Delta  \delta_Q^2 \hat \phi (\x,t)}}{\Delta \phi_0(t)} =  \frac{H^2}{2 \pi \dot \phi_0}> 1.
\end{equation} 
Equivalently, in terms of the inflaton potential and 
the first slow-roll parameter, the former condition is
\begin{equation}\label{condicionchingona}
\frac{V (\phi)}{M_P^4 \epsilon_1 (\phi)} > 1.
\end{equation} 

It is usually claimed that essentially any slow-roll potential satisfies the condition for 
eternal inflation \cite{kinney}. For instance, let us consider the potential used in 
$R^2$ inflation, also known as Starobinsky's inflation \cite{starobinsky1980},
\begin{equation}\label{R2inflation}
  V (\phi) = m^4 (1-e^{-\sqrt{2/3} \phi/M_P})^2,
\end{equation} 
which is in very good agreement with the latest observational bounds from the 
\emph{Planck} mission \cite{planck2015inflation,Martin14,infmodels}. The first slow-roll parameter for this potential is
\begin{equation}
 \epsilon_1 (\phi) = \frac{4}{3}\frac{1}{(1-e^x)^2},
\end{equation} 
where $x \equiv \sqrt{2/3} \phi/M_P$.

Therefore, the condition for eternal inflation using this type of potential is
\begin{equation}\label{eipotencial}
  \frac{m^4}{M_P^4} (1-e^{-x})^2 (1-e^x)^2 > 1.
\end{equation} 
If the inflaton is well enough in the slow-roll region $x \gg 1$, which means the field takes super-Planckian values,  then the 
condition for eternal inflation,  \eq{eipotencial}, can be satisfied because $e^{2x} \gg 1$. Furthermore, even if the energy scale is sub-Planckian $m^4 < M_P^4$, the field generically fulfills the condition for eternal inflation. In fact,  eternal inflation commonly occurs for potentials of the hilltop type \cite{kinney}. 

In the following, we will show how the traditional claims regarding eternal inflation can be modified if one takes into account the dynamical reduction of the wave function.

\section{The CSL model during inflation}\label{secCSL}

As we mentioned in the Introduction, we will use the CSL model as the particular 
model characterizing the dynamical reduction of the wave function. %Therefore, in this section, we present a brief review of the framework when considering the CSL mechanism during inflation.
The implementation of the CSL model into the picture of slow-roll inflation has been analyzed in previous works \cite{pedro,LB15,qmatter,abhishek} 
(in Refs. \cite{jmartin,hinduesS,hinduesT}  the CSL model is also considered during inflation but based on a different conceptual approach). 
Here, we present only the main features and results that will be of interest for the 
present work.

Before addressing the CSL inflationary model, we present our view regarding the relation between the spacetime description in terms of the metric and the  degrees of freedom of the inflaton (see \cite{PSS,shortcomings} for a detailed presentation). This particular view is based on the semiclassical gravity framework, which treats gravitation classically and all other fields quantum mechanically. We assume such a framework to be a valid approximation during the inflationary era, which is well after the full quantum gravity regime has ended. Note that this is a major difference between our approach and the standard one, since in the latter all scalar degrees of freedom (metric and matter fields) are quantized. On the other hand, it is not entirely settled that one should quantize the metric degrees of freedom. There are numerous arguments suggesting that the spacetime geometry might emerge from deeper, non-geometrical and fundamentally quantum mechanical degrees of freedom \cite{Em1,Em2,Em3,Em4}. Therefore, the  framework we will use is based on Einstein's semiclassical equation
\begin{equation}\label{semiclassicaleinstein}
  G_{ab} = 8 \pi G \expec{\hat T_{ab}}.
\end{equation} 

 {Let us discuss now how the dynamical reduction of the wave function fits into our general understanding of gravity and its, still incomplete, quantum mechanical characterization (see \cite{alberto,Sudarsky2014} for a more detailed explanation).  According to general relativity, gravitation reflects the structure of spacetime itself, whereas quantum theory seems to fit most easily in contexts where this structure is a given one. That is, quantum states are associated with objects that ``live'' in spacetimes. For instance, the standard Schr\"odinger equation specifies the evolution of a system; the quantum states of fields characterize the system in connection to algebras of observables associated with predetermined spacetime regions. One of many complications that become manifest in the merging of gravity and quantum mechanics, following the canonical approach, is the appearance of a timeless theory, where the recovery of fully covariant spacetime notions is nontrivial \cite{Isham1992}. Therefore, it seems natural to speculate that it is precisely in this setting where something that departs from quantum orthodoxy, as the dynamical collapse of the wave function, might find its origin. That is, the collapse process could reflect some remanent signature from a  fundamental quantum gravity regime. It is worthwhile to mention, that Penrose \cite{penrose1996} and Di\'osi, \cite{diosi1984,diosi1987,diosi2016} have suggested similar ideas regarding the connection of the collapse of the wave function and its possible relation to gravity. According to this view, it seems reasonable to conjecture that a reduction of the wave function could correspond to lingering features of a quantum gravity theory. If that is the case, the emergence of spacetime itself would be tied to the incorporation of such effective quantum description of matter fields living on the spacetime, and evolving (approximately) according to standard quantum field theory on curved spaces, with some small deviations which might include our hypothetical collapse.}

 {A specific example of this particular outlook  is  to analyze the departures from Eq. \eqref{semiclassicaleinstein} that arise when considering a single and instantaneous collapse as described by the state $| \psi (t) \ket = \Theta(t_0 - t) | 0 \ket + \Theta(t - t_0 ) | \xi \ket$ (where $\Theta(x)$ is the Heaviside function). Using that state, Eq. \eqref{semiclassicaleinstein} can be written as}
\begin{equation}
  G_{ab} = 8 \pi G  \bra 0 | \hat T_{ab} | 0 \ket + 8 \pi G \xi_{ab},
\end{equation}
 {where the object  $\xi_{ab} = \Theta (t-t_0)  ( \bra \xi | \hat T_{ab} | \xi \ket - \bra 0 | \hat T_{ab} | 0 \ket         ) $  would represent an individual stochastic step. If we extrapolating this example, then the stochastic gravity framework \cite{Calzetta1994,Verdaguer2008} might correspond to the continuous version of several discrete stochastic steps, e.g. the one given by the CSL dynamical collapse model. }
  
 {Furthermore, we accept that at the fundamental level, the spacetime would have a quantum description in terms of some unspecified variables (e.g. those of loop quantum gravity, the causal set or dynamical triangulations approaches), but during the inflationary regime the spacetime can be viewed in effectively classical terms. A useful analogy is the hydrodynamical characterization of a fluid, which at certain scales is a useful description. However, at subatomic scales one should have a radically different depiction. Einstein equations would correspond to the Navier-Stokes equations, the spacetime metric to the density and velocity fields; whereas, the subatomic characterization of the fluid constituents would correspond to the fundamental degrees of freedom of quantum gravity. Moreover, this analogy also serves to picture a situation in which departures from the semiclassical gravity approximation occur. Specifically, when a phase transition (from say liquid to gas) is taking place, one might expect not only the violation of Navier-Stokes equations but also of the hydrodynamic characterization. In our case, small violations of Eq. \eqref{semiclassicaleinstein}  might occur when the dynamical collapse of the wave function is going on, much like in the fluid analogy during a phase transition. From now on, we will assume the validity of the semiclassical Einstein equations by assuming that the dynamical process, associated with the collapse of the wave function, has reached a final (different from the initial vacuum) state. However, we should bear in mind all the open issues and subtleties that we have indicated. }

Our starting point will be the same as the standard inflationary account. We assume the initial state of the Universe characterized by the homogeneous and isotropic classical FRW spacetime and the equally symmetric Bunch-Davies vacuum. The dynamical reduction mechanism will drive the initial state of the matter field to a final state that does not need to share the symmetries of the Bunch-Davies vacuum; this mechanism acts as an effective spontaneous and stochastic quantum collapse of original the wave function. As a consequence, $\expec{\hat T_{ab}}$ will not have the symmetries of the initial state leading to a geometry, through Einstein semiclassical equation, that generically will no longer be homogeneous and isotropic.  {We would like to mention here that it might be valuable to consider the decoherence framework, plus coarse graining, since as found in Refs. \cite{markkanen2016,markkanen2017}, the time translation invariance is lost when considering said framework. Therefore, it would be interesting to investigate whether  coarse graining could potentially lead to a breakdown of the symmetries of the quantum state.   }

Focusing on the scalar metric perturbations in the longitudinal gauge, and assuming no anisotropic stress, the semiclassical equations, in Fourier space, yield
\begin{equation}\label{masterpsi}
  \Phi_{\vk}  =\sqrt{ \frac{\epsilon_1}{2}  }  \frac{ H}{ M_P  k^2} a \bra  \hat{\dphi'_{\vk}} \ket,
\end{equation} 
where $\Phi$ is the scalar metric perturbation (or the Newtonian potential in the longitudinal gauge) and $\hat \dphi$ characterizes the inhomogeneities of the inflaton.
From the former equation, we observe that in the vacuum state, the expectation value $\bra 0 | \hat {\dphi'_{\vk}} | 0 \ket$ vanishes, which implies that $\Phi_{\vk}$ is also zero for all $\vk$.  Strictly speaking, there are no perturbations and the spacetime is perfectly symmetric. It is only after the CSL mechanism has evolved the vacuum state into the post-collapse state that generically $\bra \hat{\dphi'_{\vk}} \ket \neq 0$ and, by Eq. \eqref{masterpsi}, the curvature perturbations are born. In this way, the self-induced collapse leads to a configuration in which neither the state of the field nor the spacetime is homogeneous and isotropic. 

Next, we focus on the collapse  mechanism. The CSL model is based on a modification of the Schr\"odinger equation. This alteration
induces a collapse of the wave function towards one of the possible eigenstates of an 
operator called the collapse operator, with certain rate $\lambda$. The self-induced 
collapse is due to the interaction of the system with a background noise $\mathcal{W} 
(t)$ that is characterized as a continuous-time stochastic process of the Wiener kind. 
The parameter $\lambda$ sets the strength of the collapse process; the greater the value, 
the stronger the collapse (see \cite{bassi} for a throughly review).

Since the CSL model modifies the Schr\"odinger equation, it is convenient to 
describe the theory of the inflaton in the Schr\"odinger picture, where the relevant 
objects of the theory are the wave function and the Hamiltonian.

The initial wave functional, corresponding to the quantum state given in Eq. \eqref{edoini}, is 
\begin{equation}
  \Psi_{\textrm{in}}[\phi] = \Psi_{\xi_0}[\phi_0]  
\Psi_{\textrm{BD}}[\dphi],
\end{equation} 
where $\Psi_{\xi_0}$ denotes the wave function of the homogeneous part of the field, 
which is a coherent state. Meanwhile, $\Psi_{\textrm{BD}}$ denotes the wave function 
associated to the inhomogeneous sector of the inflaton, which corresponds to the 
Bunch-Davies vacuum. 

It  will be suitable to separate the field into its real and imaginary parts; thus, in Fourier space,  the initial wave 
function is
\begin{equation}\label{psiin0}
  \Psi_{\textrm{in}}[\phi] = \Psi_{\xi_0}^{\textrm{R}} 
\Psi_{\xi_0}^{\textrm{I}}  \prod_{\vk} \Psi_{\vk}^{\textrm{R}} (\dphi_{\vk}^{\textrm{R}}) 
\Psi_{\vk}^{\textrm{I}} (\dphi_{\vk}^{\textrm{I}}),
\end{equation} 
with $\vk \neq 0$.

On the other hand, the Hamiltonian characterizing the inhomogeneous sector of the field is  $H = \frac{1}{2} 
\int d^3 k$ $ (H_{\vk}^{\textrm{R}}  + H_{\vk}^{\textrm{I}})$, with  
\begin{eqnarray}
  H_{\vk}^{\textrm{R,I}}   &=& p_{\vk}^{\textrm{R,I}} p_{\vk}^{* \textrm{R,I}} +  
k^2 y_{\vk}^{\textrm{R,I}} y_{\vk}^{* \textrm{R,I}}  \nn
&-& \frac{1}{\eta} \bigg(  y_{\vk}^\RI  p_{\vk}^{\RI*} +  y_{\vk}^{\RI*} 
p_{\vk}^{\RI}     \bigg),
\end{eqnarray} 
where we have defined the rescaled field variables 
\begin{equation}
y_{\vk}^{\textrm{R,I}} (\eta) \equiv a (\eta) \dphi_{\vk}^{\textrm{R,I}} (\eta), \qquad p_{\vk}^{\textrm{R,I}} \equiv  
(y_{\vk}^\RI)' - (a'/a) y_{\vk}^\RI. 
\end{equation} 
Additionally, we have ignored terms containing first-order slow-roll parameters since we will be mostly interested in the amplitude of the fields rather than the shape of their spectrum. Note also that $p_{\vk}^{\textrm{R,I}} = a \dphi_{\vk}^{'\:\RI}$; thus, the right hand side of Eq. \eqref{masterpsi} involves the expectation value of the momentum operator.

Promoting $y_{\vk}^\RI$ and $p_{\vk}^\RI$ to 
quantum operators, the commutations relations are $[ \hat y_{\vk}^\RI , \hat p_{\vk}^\RI ] 
=i \delta(\vk-\vk')$.

The CSL mechanism will drive the initial state vector, represented by the wave function 
$\Psi_{\textrm{in}}[\phi]$, \eq{psiin0}, towards a final state that lacks the 
original symmetries of the system. However, the zero mode part of the wave function 
(recall that $\Psi_{\xi_0}$ corresponds to a coherent state) will remain unchanged, 
this is needed for the model to be self-consistent \cite{alberto}. On the other hand, 
the collapse mechanism will change  the part of the wave function corresponding to  the 
modes $\vk \neq 0$. In our framework, these modes, together with the collapse of their wave 
function, are responsible for the birth of the primordial inhomogeneities. 

We now apply the CSL model to each (non-zero) mode of the field 
independently. The initial state vector  $ 
|\Psi_{\vk}^{\textrm{R,I}}, 
  \tau \ket $, with $\tau$ the conformal time at the beginning of inflation, will evolve 
according to
\begin{eqnarray}\label{CSLevolution}
  |\Psi_{\vk}^{\textrm{R,I}}, \eta \ket &=& \hat T \exp \bigg\{ - \int_{\tau}^{\eta} 
d\eta'   \bigg[ i \hat{H}_{\vk}^{\textrm{R,I}} \nn 
&+& \frac{1}{4 \lambda_k} (\mathcal{W}(\eta') - 2 \lambda_k^2 
\hat{p}_{\vk}^{\textrm{R,I}})^2 \bigg] \bigg\} |\Psi_{\vk}^{\textrm{R,I}}, 
  \tau \ket, 
\end{eqnarray} 
 $\hat T$ denotes the time-ordering operator. 
 
Note that we have chosen the momentum 
operator $\hat{p}_{\vk}^{\textrm{R,I}}$ as the collapse operator. This is justified 
because  the metric perturbation $\Phi$ is directly related to the expectation value of the momentum operator,  Eq. \eqref{masterpsi}. However, we could also have chosen the field variable as the collapse operator and perform a similar analysis with our conclusions  unchanged; see Ref. \cite{pedro}. The collapse parameter $\lambda_k$ depends 
on each mode. From dimensional analysis, one finds that $\lambda_k$ is dimensionless. 
Additionally,  $\lambda_k$ sets the strength of the collapse, and since in principle all 
non-zero modes are subjected to it, we expect that $\lambda_k \gg 1$ for all $k$.  

 {In addition, we want to highlight some open issues regarding the CSL inflationary model.   First, the CSL model is a non-relativistic model; hence, in order to use it in the cosmological setting, it is required to introduce  suitable modifications; see \cite{elias2017} for a detailed discussion. However, because of the perturbative nature of our analysis, such modifications will not be of importance  in the present work. A second related issue is that, as is known \cite{bassi2005}, the CSL model in general violates energy conservation, although the energy increase is minimal (for example a particle of mass $m = 10^{-23}$ g, would take $10^{10}$ years for an energy increase of $10^{-8}$ eV \cite{bassi}). Nevertheless, even if the energy violation can be neglected, at the fundamental level a more realistic model should remove this issue. In the case of the CSL inflationary model, the energy violation issue could lead to divergences in the energy-momentum tensor (see a recent work in which these violations in the energy momentum tensor are explicitly computed \cite{benito} and shown to be small). Third, let us recall that the amplification mechanism of the CSL model is the idea that the collapses must be rare for microscopic systems, but the effect of the collapses must increase when several particles are hold together forming a macroscopic system. This amplification mechanism is included in the CSL inflationary model in the $k$ dependence of the $\lambda_k$ parameter. Nonetheless, the explicit $k$ dependence, as well as the choice of the collapse operator, is purely phenomenological at this point. In our opinion, this third open issue is the most pressing one at the moment for the collapse models in the inflationary setting. We will say more about this at the end of the present section. }

Returning to the CSL model during inflation, characterized by Eq. \eqref{CSLevolution}, we note that the initial wave function of each non-zero mode of the field is represented by a 
Gaussian centered at zero with certain spread. We expect that the CSL mechanism will not 
change the Gaussian nature of the evolved state (clearly, the spread and the peak of the 
Gaussian will not be the same as the original). Also, since we have selected $\hat 
p_{\vk}^\RI$ as the collapse operator, it will be suitable to work in the momentum 
representation; thus, the wave function is characterized by
\begin{eqnarray}\label{psicsl}
  & &  \bra p_{\vk}^\RI | \Psi_{\vk}^\RI , \eta \ket    =   \Psi_{\vk}^\RI(\eta,p_{\vk}^\RI)   = \nn
 & & \exp\left[-A_k(\eta) (p_{\vk}^\RI)^2 + B_k^\RI (\eta) 
p_{\vk}^\RI + C_k^\RI (\eta) \right]. \nn
\end{eqnarray} 
The dynamical CSL evolution of the wave function follows \eq{CSLevolution}, 
provided by the initial conditions $A_k(\tau) = (2k)^{-1}, B_k^\RI (\tau)=0, C_k^\RI 
(\tau) =0$, corresponding to the Bunch-Davies vacuum. In fact, for the purpose of the 
present paper, we will be only interested in $A_k(\eta)$. The evolution equation for such a
quantity is
\begin{equation}
A_k' = \frac{i}{2} + \lambda_k - \frac{2}{\eta} A_k  - 2i k^2 A_k^2.
\end{equation} 
This equation can be solved by performing a change of variable $A_k (\eta) \equiv f'(\eta)/[2ik^2 f(\eta)]$, which results in a Bessel differential equation for $f$. After solving such an equation, and returning to the original variable $A_k$, we obtain
\begin{equation}\label{Aketa}
  A_k (\eta) = \frac{q}{2ik^2} \left[   \frac{J_{3/2} (-q\eta) + e^{-i \pi/2} J_{-3/2} (-q\eta)  }{J_{1/2} (-q\eta) - e^{-i \pi/2} J_{-1/2} (-q\eta) }      \right],
\end{equation} 
where $q^2 \equiv k^2 (1-2 i \lambda_k)$, and $J_n$ represents a Bessel function of order $n$.

One of the reasons we are interested in the quantity $A_k(\eta)$ is because it is related to the prediction for the scalar power spectrum within the CSL inflationary model. Recall that the dimensionless  power spectrum in Fourier space is defined as
\begin{equation}
 \avg{\Phi_{\vk} \Phi_{\vk'}^*  } \equiv \frac{2\pi^2}{k^3} \mP (k) \delta(\vk-\vk'),
\end{equation} 
where $\avg{\cdot}$ denotes an ensemble average over possible realizations of the stochastic field $\Phi$. In our approach, each realization of $\Phi$ corresponds to a particular post-collapse state; the stochasticity of the post-collapse state is generated  from the noise function $\mathcal{W}$.  
Equation \eqref{masterpsi} implies that 
\begin{eqnarray}
 \avg{\Phi_{\vk} \Phi_{\vk'}^*  } &\propto&  \avg{ \expec{\hat p_{\vk}} \expec{\hat p_{\vk'}}^*     } \nn
 &=&  (\avg{ \expec{ \hat p_{\vk}^{\textrm{R}}  }^2  } + \avg{ \expec{ \hat p_{\vk}^{\textrm{I}}  }^2  } )\delta(\vk-\vk'),
 \end{eqnarray} 
 as a matter of fact, $ \avg{ \expec{ \hat p_{\vk}^{\textrm{R}}  }^2  } = \avg{ \expec{ \hat p_{\vk}^{\textrm{I}}  }^2  } $. In can be shown that \cite{pedro}
\begin{equation}
  \avg{\expec{\hat p_{\vk}^\RI}^2} = \avg{\expec{\hat p_{\vk}^{\RI \: 2}}} - \frac{1}{4 \textrm{Re}[A_k (\eta)]},
\end{equation} 
The predicted scalar power spectrum, within the CSL framework,  is explicitly calculated in \cite{pedro}. The result is 
\begin{equation}\label{PSCSL}
  \mP(k) \simeq  \frac{V}{M_P^4 \epsilon_1} \lambda_k  k |\tau|.
\end{equation} 
which is scale invariant if $\lambda_k = \lambda/k$ (recall that $\lambda$ is the universal CSL parameter). 
% 
% Note once again that if $\lambda=0$, i.e. there is no collapse and the vacuum evolves according to the Schr\"odinger equation, then  $\mP(k)=0$, which means that there are no inhomogeneities at any scale; this is consistent with our conceptual approach.

 {At this point we want to discuss some important aspects  that led to the latter result. First, we can observe that the explicit dependence on $k$ of the $\lambda_k$ parameter is purely phenomenological, and at the time of writing we cannot provide a more fundamental justification. Additionally, in order to obtain that result one had to fix a particular gauge (the longitudinal gauge) even before the quantization process. This gauge fixing is inevitable, since the use of the semiclassical gravity framework implies that matter and geometry degrees of freedom should be treated differently (the former are quantum mechanical while the latter are always classical).  Also, the choice of  gauge  implies that the time  coordinate is attached  to  some specific  slicing of  the  perturbed spacetime, and thus our identification of  the corresponding  hypersurfaces (those of constant time)  as the  ones  associated with the  occurrence of collapses--something deemed as an actual physical  change--turns  what is normally  a simple  choice of  gauge  into a choice of   the distinguished  hypersurfaces,  tied to the putative physical process behind  the collapse. This aspect of the collapse process was also found in Ref. \cite{finl}. Naturally this is an issue that leads to  tensions  with the expected   general covariance of a fundamental theory, a  problem that afflicts  all known  collapse  models, which in the non-gravitational   settings becomes the issue   of compatibility with  Lorentz  invariance of the proposals.  We must acknowledge that this generic problem  of  collapse  models is indeed  an  open issue for the present approach. One  would expect that its  resolution  would be tied to the  uncovering of the  actual  physics  behind  what we treat here as  the  collapse of  the  wave function (which we  view as a merely an effective description). As has been argued in related  works, and  in following ideas originally  exposed  by Penrose \cite{penrose1996},   we hold that the physics  that  lies behind  all  this,  ties the  quantum  treatment of gravitation  with the foundational issues  afflicting quantum theory in general, and in particular those  with connection  to the quantum measurement problem. }

Finally, it will be convenient to write Eq. \eqref{PSCSL}  re-expressing the potential as $V(\phi) = m^4 V_0(\phi/\mu)$ (that is $V_0$ is dimensionless). The predicted spectrum is then
\begin{equation}\label{PSCSL2}
  \mP(k) \simeq \frac{m^4}{M_P^4} \frac{V_0^* }{\epsilon_1^*} \lambda |\tau|,
\end{equation} 
where $V_0^*$ and $\epsilon_1^*$ indicate that such functions are being evaluated at the time $\eta_*$  when the amplitude of the curvature perturbation is equal to the temperature of the anisotropies observed in the CMB, $(\delta T/T)^2_{\textrm{CMB}} \simeq 10^{-9}$.

\section{Modification of the standard picture}\label{mod}

Having established the main features of the CSL inflationary model, we proceed to show how this model modifies the standard results regarding eternal inflation. 

We consider first the expectation value of the inflaton in the state that resulted from the CSL mechanism, i.e. 
\begin{eqnarray}\label{expecCSL}
\bra \Psi_{\textrm{CSL}} | \hat \phi (\x,\eta) |   \Psi_{\textrm{CSL}} \ket &=& \bra \Psi_{\textrm{CSL}} | \hat \phi_0 (\eta) |   \Psi_{\textrm{CSL}} \ket  \nn 
&+& \varepsilon \bra \Psi_{\textrm{CSL}} | \hat \dphi (\x,\eta) |   \Psi_{\textrm{CSL}} \ket.
\end{eqnarray} 
As we mentioned in the previous section, the CSL mechanism only affects the non-zero modes of the field; therefore, the initial state of the zero mode, which is a coherent state, remains unchanged. On the other hand, the expectation value of $\hat \dphi$ can be characterized as 
\begin{equation}
 \bra \Psi_{\textrm{CSL}} | \hat \dphi (\x,\eta) |   \Psi_{\textrm{CSL}} \ket \equiv \dphi(\x,\eta).  
\end{equation} 
Furthermore, since the CSL mechanism is based on a stochastic process in which each realization of the noise $\mathcal{W}_\alpha$, with $\alpha$ denoting the particular realization, corresponds to a different state $ |   \Psi_{\textrm{CSL}} \ket$,  the field $\dphi(\x,\eta)$ is a classical stochastic field. 

The expectation value given in \eq{expecCSL} is then
\begin{equation}\label{expecCSL2}
 \bra \Psi_{\textrm{CSL}} | \hat \phi (\x,\eta) |   \Psi_{\textrm{CSL}} \ket  = \phi_0 (\eta) + \varepsilon \dphi(\x,\eta) \equiv \phi (\x,\eta).
\end{equation}

Switching back to cosmic time,  the variation $\qquad  $ $\Delta \bra \Psi_{\textrm{CSL}} | \hat \phi (\x,t) |   \Psi_{\textrm{CSL}} \ket$ over one Hubble time $\Delta t = H^{-1}$, using Eq. \eqref{expecCSL2}, is
\begin{eqnarray}
 \Delta \bra \Psi_{\textrm{CSL}} | \hat \phi (\x,t) |   \Psi_{\textrm{CSL}} \ket  &=& [ \dot \phi_0(t) \nn
 &+& \varepsilon \dot \dphi (\x,t)] \Delta t = \Delta \phi. 
\end{eqnarray} 
Moreover, $\varepsilon \dot \dphi$ must be smaller than $\dot \phi_0$ in order to maintain the consistency of the perturbative analysis. 
% , hence, $\Delta \phi \simeq \dot \phi_0 \Delta t = \Delta \phi_0$. 
Consequently,  the variation of the expectation value of the field over one Hubble time is 
\begin{equation}\label{Deltaph0}
 \Delta \phi_0 =   \frac{ \dot \phi_0 }{H}.
\end{equation}

The previous analysis contrasts with the corresponding one from the  standard picture. As we saw in section \ref{standard}, the expectation value of the field  $\hat \phi (\x,\eta)$ is taken with respect to the initial state,  Eq. \eqref{edoini}, during the whole inflationary regime; henceforth, $\bra \textrm{in} | \hat \phi (\x,\eta)| \textrm {in}  \ket = \phi_0(\eta)$. In other words, in the standard approach, if one wishes to characterize classically the field $\phi (\x,\eta)$ from the  quantum fields $\hat \phi_0$, $\hat \dphi$ and using only the $|\textrm{in} \ket$ state, then there is an issue because the expectation value of $\hat \phi$ can only provide homogeneous fields. In order to overcome such an issue,  the traditional posture postulates the association $\dphi (\x,\eta) \sim \bra \hat \dphi (\x,\eta)^2 \ket^{1/2}$, which, in principle, it is not clear where the stochasticity of the field $\dphi (\x,\eta)$ comes from. In fact, after renormalization \cite{vilenkin1982}, the quantum expectation value $\bra \hat \dphi (\x,\eta)^2 \ket$ leads to a result that is independent of $\x$, and thus it is homogeneous. 

Conversely, in our approach, the classical nature of the fields $\phi_0$ and $\dphi$, as well as their stochastic origin, are clearly identified in Eqs.  \eqref{expecCSL} and \eqref{expecCSL2}.

We now focus on the quantum uncertainty of the field $\hat \phi(\x,\eta)$ using the CSL evolved state,  i.e.
\begin{eqnarray}
 \delta_Q^2 \hat \phi (\x,\eta) &\equiv& \bra \Psi_{\textrm{CSL}} | \hat \phi (\x,\eta)^2 |   \Psi_{\textrm{CSL}} \ket \nn
 &-& ( \bra \Psi_{\textrm{CSL}} | \hat \phi (\x,\eta) |   \Psi_{\textrm{CSL}} \ket )^2.
\end{eqnarray} 
In discrete Fourier space, the latter expression is
\begin{eqnarray}\label{x}
  \delta_Q^2 \hat \phi (\x,\eta) &=&  \frac{\varepsilon^2}{L^6} \sum_{\vk,\vk'} \bigg( \bra \hat \dphi_{\vk}^{\textrm{R}} \hat \dphi_{\vk'}^{\textrm{R}} \ket -  \bra \hat \dphi_{\vk}^{\textrm{I}} \hat \dphi_{\vk'}^{\textrm{I}} \ket \nn
  &-&  \bra \hat \dphi_{\vk}^{\textrm{R}}  \ket \bra \hat \dphi_{\vk'}^{\textrm{R}}  \ket + \bra \hat \dphi_{\vk}^{\textrm{I}}  \ket \bra \hat \dphi_{\vk'}^{\textrm{I}}  \ket \bigg) e^{i \x \cdot (\vk+\vk')} . \nn
\end{eqnarray} 
Note that only when $\vk = -\vk'$, one obtains a non-vanishing result. Using the wave function given in \eq{psicsl}, and taking the continuum limit $L \to \infty$, we find 
\begin{equation}\label{incertCSL0}
  \delta_Q^2 \hat \phi (\x,\eta) =  \frac{1}{a^2 \pi^2} \int dk \: k^2 \frac{| A_k(\eta)|^2}{ \textrm{Re} [A_k(\eta)] }.
\end{equation} 
We have omitted the $\varepsilon^2$ factor as the only role played by $\varepsilon$ is to ensure that the scalar perturbations are small compared to the background. Equation \eqref{incertCSL0} implies that the uncertainty $  \delta_Q^2 \hat \phi (\x,\eta) $ is essentially equal to the sum (integral) of the uncertainty of each mode of the field, that is, 
$\delta_Q^2 \hat \phi_{\vk} (\eta) \simeq {| A_k(\eta)|^2}/{a^2 \textrm{Re} [A_k(\eta)] }  $.

An estimated value for the variation of  $ \delta_Q^2  \hat \phi $ over one Hubble time can be found following the same procedure as in the standard approach. Namely, we introduce an UV cut-off $k_H (\eta)$ that selects only the super-Hubble modes. Using Eqs. \eqref{Aketa} and \eqref{incertCSL0},  the variation  $\Delta \delta_Q^2  \hat \phi $ over $\Delta t = H^{-1}$ is 
\begin{eqnarray}
  \Delta \delta_Q^2 \hat \phi (\x,t) &=&   \frac{\Delta t  H^2}{2 \pi^2}  \frac{d\eta}{dt}  \nn
  &\times& \frac{d}{d\eta} \bigg\{ \int_0^{k_H(\eta)}\frac{dk}{k} \:  \frac{|1-i k\eta \zeta_k e^{i\theta_k} |^2}{\textrm{Re}[\zeta_k e^{i \theta_k}]} \bigg\}, \nn
\end{eqnarray} 
where we have defined $\zeta_k e^{i \theta_k} \equiv \sqrt{1-2i \lambda_k}$. Furthermore, since $k_H(\eta) = -1/\eta$, and $d\eta/dt = a^{-1} = -H\eta$, we obtain
\begin{equation}\label{incertCSL}
  \Delta \delta_Q^2 \hat \phi (\x,t) =   \Delta t  \frac{  H^3}{2 \pi^2}  \frac{2 }{ \sqrt{ 1+ (1+4 \lambda_{H}^2)^{1/2}   }  },
\end{equation} 
with
\begin{equation}
  \lambda_{H} \equiv \lambda|\tau|e^{-\Delta N}, 
\end{equation} 
and $\Delta N$ the number of e-folds from the beginning of inflation to some time $\eta$.

From Eq. \eqref{incertCSL}, we observe that if there is no collapse, i.e $\lambda=0$,  which implies that the state is the Bunch-Davies vacuum, then we recover the estimate of the standard treatment; see Eq. \eqref{varphiquant}.

Equation \eqref{incertCSL}, also serves to illustrate the effect of the collapse. As long as $\lambda_H  \gg 1$, the uncertainty of the field $\delta_Q^2 \hat \phi$ will start to decrease with respect to the uncertainty in the vacuum state, which is expected from a reduction of the wave function.  In fact, let us consider a super-Hubble mode $k^+$, that is, $k^+ |\eta| < 1$. If $\lambda_H \gg 1$ then $\lambda/k^+ \gg  1$; henceforth, the collapse affects the modes whose physical wavelength is greater than the Hubble radius during inflation, which are also the ones that originate the seeds of the observed structure in the Universe. Contrarily, for a sub-Hubble mode $k^-$, i.e. $k^- |\eta| > 1$, if $\lambda_H  \ll 1$ then $\lambda/k^- \ll 1$; this implies that the uncertainty depicted in Eq. \eqref{incertCSL} is essentially the same as the one obtained in the vacuum state. Thus, in the CSL reduction model, the collapse affects the super-Hubble modes, during inflation, more than the  sub-Hubble modes. 

% 
% In fact, since $\lambda_{H} = \lambda/k_H$, with $\lambda$ the universal CSL parameter, then $\lambda$ must be such that $\lambda \gg k_H$ during the whole inflationary phase, to ensure that the collapse 
% is strong enough, i.e. it affects the modes whose physical wavelength is greater than the Hubble radius during inflation, which are also the ones that give birth to the observed structure in the Universe. 

The main information contained in Eq. \eqref{incertCSL} is that the CSL evolution strongly localizes the wave function of the field if $\lambda_H \gg 1$. The localization is achieved around the expectation value of the field, i.e. $\phi_0 + \dphi$.  Hence, the wave function of the inflaton is effectively centered around $\phi_0$, which obeys its classical equation of motion. Given that $\dphi < \phi_0$, the evolution of $\phi_0$ continues essentially undisturbed. 

Next, we consider whether  the condition for eternal inflation \eqref{condicionchingona0}  is fulfilled within the CSL inflationary framework. The condition for eternal inflation occurs when the variation of the width of the wave function amplitude is bigger than the variation of its peak over a Hubble time.

The ratio of the variation of the quantum uncertainty with respect to the variation of the expectation value of the field, over a Hubble time, can be obtained from Eqs. \eqref{Deltaph0} and \eqref{incertCSL}. This ratio is
\begin{equation}\label{chingona}
  \frac{\sqrt{ \Delta \delta_Q^2 \hat \phi (\x,t) }}{\Delta \phi_0(t)} = \frac{H^2}{2 \pi \dot \phi_0}  \frac{2 }{ \sqrt{ 1+ (1+4 \lambda_{H}^2)^{1/2}   }  }.
\end{equation} 
Equation \eqref{chingona}, is the main result of this section. 

Consequently, if $\lambda_{H} \gg 1$ then, in terms of the potential and the first slow-roll parameter, expression \eqref{chingona} is
\begin{equation}\label{chingona2}
  \frac{\sqrt{ \Delta \delta_Q^2 \hat \phi (\x,t) }}{\Delta \phi_0 (t)} \simeq \left( \frac{V}{M_P^4 \epsilon_1} \frac{1}{\lambda_{H}} \right)^{1/2}.
\end{equation} 
As we will show next, using a particular potential, the assumption $\lambda_{H} \gg 1$ implies that 
\begin{equation}\label{violacion}
  \frac{\sqrt{ \Delta \delta_Q^2 \hat \phi (\x,t)}}{\Delta \phi_0 (t)} < 1,
\end{equation} 
which is the opposite of the condition for eternal inflation to occur. 

On the other hand,  the prediction for the scalar power spectrum is  directly proportional to $\lambda$ as illustrated in Eq. \eqref{PSCSL2}. If $\lambda_H \gg 1$ then we must adjust $m$ such that the overall amplitude of the power spectrum fits the observational data, i.e. it must be consistent with $\mP(k) \simeq 10^{-9}$. In particular, since $\lambda |\tau|$ is a large number, the factor $m^4/M_P^4$ must be quite small in order to satisfy 
\begin{equation}
  \frac{m^4}{M_P^4} \frac{ V_0^*}{\epsilon_1^*} \lambda |\tau| \simeq 10^{-9}.
\end{equation}
Moreover, given the smallness of $m^4/M_P^4$, Eq. \eqref{chingona2} is ensured to be quite small (i.e. $\ll 1$).

To further illustrate our picture, let us consider once again the $R^2$ inflationary model; the potential is given in Eq. \eqref{R2inflation}. In this model, the number of e-folds from the beginning of inflation to some time $\eta$, is \cite{infmodels}
\begin{equation}\label{Nfolds}
  \Delta N \simeq \frac{3}{4} \left( e^{x_\tau} - e^x \right),
\end{equation} 
recall that $x \equiv \sqrt{2/3} \phi/M_P$, so $x_\tau$ denotes the value of the field at the conformal time $\tau$ corresponding to the beginning of inflation.  
From Eq. \eqref{chingona2}, we obtain
 \begin{equation}
   \frac{\sqrt{ \Delta \delta_Q^2 \hat \phi (\x,t) }}{\Delta \phi_0 (t)} \simeq \frac{m^2}{M_P^2} \frac{e^{x+\Delta N/2}}{({\lambda \tau})^{1/2}}.
\end{equation} 
But given Eq. \eqref{Nfolds} it is clear that $\Delta N \gg x$, hence
 \begin{equation}\label{chingona4}
  \frac{\sqrt{ \Delta \delta_Q^2 \hat \phi (\x,t) }}{\Delta \phi_0 (t)} \simeq \frac{m^2}{M_P^2} \frac{e^{\Delta N/2}}{({\lambda \tau})^{1/2}} = \frac{m^2}{M_P^2}  \frac{1}{\lambda_{H}^{1/2}}
\end{equation} 
Therefore, the assumption $\lambda_{H} \gg 1$, and considering sub-Planckian energy scales, implies that the condition for eternal inflation is bypassed, Eq. \eqref{violacion}.

In order to fit the observed amplitude of the spectrum, the mass of the inflaton must be 
\begin{equation}\label{masa}
  \frac{m^4}{M_P^4} \simeq \frac{e^{-2x_*}}{\lambda |\tau| }  10^{-8}  ,
\end{equation} 
where $x_*$ refers to the value of the field at the time $\eta_*$.

\section{Discussion and conclusions}\label{disc}

In this paper, we have considered an additional element to the standard picture of eternal inflation  {of the slow-roll/chaotic type.} This element is the dynamical reduction of the wave function. The initial wave function of the field $\hat \phi$ is characterized by a Gaussian, centered around $\phi_0$, with certain spread $\delta_Q \hat \phi$. The value of $\phi_0$ is associated with the expectation value of $\hat \phi$.

The traditional condition for eternal inflation is
\begin{equation}\label{final1}
  \frac{ \sqrt{ \Delta \delta^2_Q \hat \phi(\x,t)}}{\Delta \phi_0} \simeq \frac{H^2}{2 \pi  \dot \phi_0} > 1. 
\end{equation} 
On the other hand, when we apply the self-induced collapse of the wave function, provided by the CSL mechanism, to the inflationary Universe we find that it is possible to overturn the condition for eternal inflation. In particular, we obtain 
\begin{equation}\label{final2}
  \frac{ \sqrt{ \Delta \delta^2_Q \hat \phi(\x,t)}}{\Delta \phi_0} \simeq \frac{H^2}{2 \pi  \dot \phi_0} \frac{2}{\sqrt{1 + (1+ 4 \lambda_{H}^2)^{1/2}} } < 1, 
\end{equation} 
as long as $\lambda_{H} \equiv \lambda|\tau| e^{-\Delta N} \gg 1$ holds.  The parameter $\lambda$ is the universal CSL collapse parameter, $\tau$ is the conformal time at the beginning of inflation, and $\Delta N$ the number of e-folds from $\tau$ to some time $\eta$ during inflation.

Equation \eqref{final1} indicates that, during one Hubble time, the variation of the quantum uncertainty of the field becomes larger than the variation of the homogeneous field $\phi_0$. In other words, in the standard setting, one claims that the quantum uncertainties become actual dynamical fluctuations that dominate over the classical dynamics of $\phi_0$.  Therefore, in the standard scenario, as the Universe inflates, the spread of the wave packet, corresponding to the wave function amplitude, increases more rapidly than the variation of its peak, failing to localize the center of the wave packet. And since the dynamics of the background is driven by $\phi_0$, one cannot assure that inflation will end because the peak of the wave packet, which follows the classical trajectory, has been diluted inside its width. Conversely, Eq. \eqref{final2} shows that if the collapse is strong enough, i.e. $\lambda_H \gg 1$, then the width of the wave function will decrease, localizing itself effectively around $\phi_0$, which will drive inflation to an end.  

In the traditional picture, the failure to localize $\phi_0$ along the wave function is sometimes used to claim that the field is as likely to roll up as it is to roll down the potential, leading to an endless cycle of inflation for different regions of the Universe since the value of the field, in distinct regions, varies according to the spread of the Gaussian characterizing the initial wave function of the field. In our model, the CSL reduction mechanism strongly localizes the wave function around the expectation value of the field $\bra \hat \phi \ket \simeq \phi_0$ in any sector of the potential.  Once properly localized, the field $\phi_0$ will evolve essentially unperturbed according to its classical equation of motion; this occurs in any region of the Universe indistinctly. 
% Thus, once the collapse has localized the Gaussian around $\phi_0$, that region of the Universe will no longer undergo into another cycle of inflation.   

Another important difference between the standard approach and the CSL inflationary model involves the prediction for the power spectrum. In the traditional setting, the theoretical prediction for the power spectrum is of the same structure as the left-hand side of Eq. \eqref{final1}. On the other hand, in our approach, the predicted expression for the power spectrum is $\sim H^2 \lambda |\tau| / \dot \phi$, which is actually a different expression from the ratio 
$  { [ \Delta \delta^2_Q \hat \phi(\x,t)]^{1/2}}/{\Delta \phi_0}$ that results from the CSL model; see Eq. \eqref{final2}. The fact that two latter expressions are different, together with the additional parameter $\lambda$, provides our model with more freedom to achieve the observed amplitude of the spectrum and to evade the condition for eternal inflation. Concretely, the CSL inflationary model contains two main parameters: the collapse parameter $\lambda$ and the overall energy scale of inflation $m^4$. As seen from Eq. \eqref{final2}, the condition for eternal inflation can be avoided if $\lambda |\tau| e^{-\Delta N} \gg 1$; additionally, the energy scale of inflation $m^4$ has to be adjusted to fit the observed amplitude of the spectrum. In the standard inflationary picture, the parameter $m^4$ is also fixed by the amplitude of the spectrum. However, once this parameter is fixed there is no other parameter that helps to bypass condition for eternal inflation resulting in a generic feature of most inflationary models. 

An additional aspect worth considering is the implication when $\lambda_H \ll 1 $. This situation virtually entails that no collapse has occurred for a particular set of modes, namely, the sub-Hubble modes. Moreover, the uncertainty of the field $\hat \phi (\x,t)$ is essentially a sum (integral) of the uncertainties of all mode operators $\hat \phi_{\vk} (t)$. In particular, $\delta_Q^2 \hat \phi (\x,t)$ is a sum of two kinds of quantum uncertainties. One kind is associated to the sub-Hubble modes, which we know are roughly equal to the uncertainty of the vacuum state. The other type encloses the uncertainties corresponding to the super-Hubble modes, which are decreasing due to the spontaneous collapse process. Henceforth, in the CSL inflationary model, the localization of $\hat \phi (\x,t)$ is not completely perfect because the uncertainty of the sub-Hubble modes is bigger than the uncertainty of the super-Hubble modes.  We think  this is a trait inherited by considering the collapse acting on each mode of the field. We expect that a covariant reduction mechanism of the wave function, that acts on the field variables $\hat \phi (\x,t)$ and/or its conjugated momentum field, will alter this picture leading  to a sharp localization of $\hat \phi (\x,t)$, or, equivalently, to a collapse that affects all the field modes equally. However, even if the localization of $\hat \phi$ provided by the CSL model is not perfect, the field expectation value consists of the zero mode plus $\bra \hat \dphi \ket$, which is always smaller than $\phi_0$. Thus the zero mode does not suffer from  large disturbances.

In summary, we have considered whether or not the standard arguments, regarding eternal inflation, are modified by considering the self-induced collapse of the wave function. Although there might be some issues related to the initial conditions for inflation \cite{ijjas}; once inflation has started in a region of the spacetime, the dynamical reduction of the wave function will generate the primordial inhomogeneities corresponding to all the observed structure in the Universe. After the primordial perturbations are born, the wave function of the field will be narrowly centered around $\phi_0$,  disrupting the condition for eternal inflation.  {Note that the analysis shown in the present work, which considered the self-induced collapse proposal, only focused on the eternal inflation process of the slow-roll/chaotic type. Henceforth, our analysis would modify the usual conclusions regarding a never-ending inflationary phase and all its ramifications concerning the multiverse, only to those kind of inflationary models. On the other hand, we left for future work the analysis concerning the eternal inflation mechanism based on false vacuum decay \cite{vilenkin1983,guth2007}, which many consider the essential concept leading to the multiverse landscape. }

\begin{acknowledgements}

The author acknowledges having useful discussions with G. R. Bengochea and D. Sudarsky, and financial support from CONICET, Argentina. 

\end{acknowledgements}

\bibliography{bibliografia0}
\bibliographystyle{apsrev}

\end{document}